\documentclass[prd,aps,showpacs,preprintnumbers,amssymb]{revtex4}

\usepackage{color}
\usepackage{epsf}

\def\e3p{$\eta \rightarrow 3 \pi$}

\begin{document}
\title{%
\hfill{\normalsize\vbox{%
\hbox{}
 }}\\
{Note on the standard model as an effective theory}}

\author{Amir H. Fariborz
$^{\it \bf a}$~\footnote[1]{Email:
 fariboa@sunyit.edu}}

\author{Renata Jora
$^{\it \bf b}$~\footnote[2]{Email:
 rjora@theory.nipne.ro}}

\affiliation{$^{\bf \it a}$ Department of Matemathics/Physics, SUNY Polytechnic Institute, Utica, NY 13502, USA}
\affiliation{$^{\bf \it b}$ National Institute of Physics and Nuclear Engineering PO Box MG-6, Bucharest-Magurele, Romania}

\date{\today}

\begin{abstract}
We present a criterion of consistency derived from the analogy between the partition function of a quantum field theory and that of a statistical system.  Based on this we examine a new class of higher dimension operators that might act in the standard model.
\end{abstract}
\maketitle

\section{Introduction}
The standard model of elementary particles \cite{Glashow}-\cite{Veltman} is a remarkably successful theory which up to now has passed almost all experimental benchmarks. However from the purely theoretical point of view has failed to provide explanations for a string of problems ranging from connection with cosmology to gauge hierarchy or fermion masses. It is now widely accepted that although  very good as an effective theory at the electroweak scale the standard model needs additional particles and interactions or simply symmetries to be an acceptable theory up to the Planck scale. If additional particles are introduced these need to be integrated out and if the symmetries are extended these need to be spontaneously or explicitly broken for the standard model to survive at the weak scale as it is. In both cases the presence of new particles and symmetries would be perceived at  lower scales as supplementary higher order operators that respect the standard model content and symmetries. In the present work is this latter approach that we shall adopt: Thus we study what kind of operators of dimension six, eight or higher one might need to introduce in the standard model for this to satisfy a criterion of consistency that we will present and discuss in the next section. The type of operators we examine do not exhaust the list of higher dimension operators that might exist in the standard model, so they are necessary but by no mean sufficient.

%By applying this we shall find a list of operators thatw e claim must exist in the standard model as an effective theory.  However this list will not exhaust the number of possible operators that might contribute. Thus our %list is necessary but by no mean sufficient.

\section{A consistency criterion for the standard model}

The standard model contains three generations of fermions, each with $15$ left handed and/or right handed fermion states and $12$ gauge bosons corresponding to the group $U(1)\times SU(2)_L\times SU(3)_c$. At scales larger than the electroweak scale the symmetry group is unbroken and all particles except the Higgs boson which counts in total $4$ real degrees of freedom are massless. We are interested to study the full partition function in the standard path integral approach. The unaltered  partition function  of the standard  model contains in fact a full dependence in the Fourier space on all the momenta and kinetic terms of the particles involved .  It is true that the partition function is usually normalized to $1$ or assumed to be $1$ because in all physical processes or correlators one divides over the full partition function  and thus consider its true value irrelevant.  In what follows we try to investigate the full dependence on the momenta and its consequences with the hope that this might bring some useful insight into the standard model.

We start by writing the exact partition function for the standard model:
\begin{eqnarray}
Z={\rm const}\prod_p \det(\gamma^{\mu}p_{\mu})^{6N}\det(\gamma^{\mu}p_{\mu})^3\det(\bar{\sigma}^{\mu}p_{\mu})^3\frac{1}{(p^2)^{\frac{12d}{2}}(p^2-m^2)^4}(p^2)^{12}\exp[\sum_i V_i].
\label{fullpart718}
\end{eqnarray}
Here the first factor corresponds to the six flavors of quarks where $N=3$ is the number of colors; the second to the three charged leptons whereas the fourth to the neutrinos. In the denominator is the gauge boson contributions and that of the Higgs bosons; the fifth factor corresponds to the ghosts. The last factor contains the contribution of vacuum bubble diagrams. Note that we did not drop any of the kinetic term contribution that appear in the standard model Lagrangian. One can consider a simplification of the mass terms in Eq. (\ref{fullpart718}) as follows:
\begin{eqnarray}
&&\prod_p\frac{1}{p^2-m^2}=\prod_p\frac{1}{p^2}(1-\frac{m^2}{p^2})=
\nonumber\\
&&\prod_p \frac{1}{p^2}\exp[\sum_p \ln[1-\frac{m^2}{p^2}]]=\prod_p \frac{1}{p^2}\exp[\sum_p[-\frac{m^2}{p^2}+...]].
\label{res5546}
\end{eqnarray}
Thus the first product remains as such in the expression for the partition function whereas the exponent can be added to $\sum_i V_i$ thus contributing to the vaccuum energy.

One can compute the factors in Eq. (\ref{fullpart718}) and simplify the expression to:
\begin{eqnarray}
Z={\rm const}\prod_p (p^2)^{31}\exp[\sum_i V_i].
\label{res55466}
\end{eqnarray}

In order to evaluate  the constant in front one can consider a space time lattice in the euclidean space where $VT$ is the total volume ($V$ is the space volune and $T$ is the time interval). Since in the path integral approach each $p^2$ comes with a factor of $\frac{1}{VT}$ (see \cite{Peskin} for the relevant information) and also factors of $i$ contribute the exact expression for the partition function becomes:
\begin{eqnarray}
Z=\prod_p \left(\frac{p^2}{VT}\right)^{31}\exp[\sum_i V_i].
\label{ex43556}
\end{eqnarray}

We need also to determine a suitable expression for the vacuum bubbles $\sum_iV_i$. For that consider a simple vacuum bubble diagram with $k $ propagators. The expression for it up to some couplings in front is:
\begin{eqnarray}
\int d^4p_1... d^4p_k\frac{1}{p_1^2p_2^2..p_k^2}\delta(p_1+...+p_k).
\label{res6657}
\end{eqnarray}
For a lattice space time with volume $VT$ this expression becomes:
\begin{eqnarray}
\frac{1}{(VT)^{k-1}}\sum_{p_1}..\sum_{p_k}\frac{VT}{p_1^2}..\frac{VT}{(p_1+p_2+...p_{k-1})^2},
\label{expr7768}
\end{eqnarray}
because each integral comes with factor of $\frac{1}{VT}$ and each propagator with a factor of $VT$. It is clear then that there is factor of $VT$ in front of each expression of the type $V_i$. Based on these reasons but also from more general arguments (see \cite{Peskin}) one concludes that:
\begin{eqnarray}
\exp[\sum V_i]=\exp[-E_02T]
\label{rez5564}
\end{eqnarray}
where $E_0$ is the energy of the vacuum and $T$ is the time interval in the euclidean space.

Now we will use the a naive  equivalence that exists between statistical mechanics and QFT which states that one can apply the rules of statistical mechanics to QFT provided that
one considers ${\it temperature}=\frac{1}{\hbar T}$. We shall also set the Boltzmann factor $k_B=1$ and $\hbar=1$. For a lattice that counts the momenta up to $k_N$ where $N$ is a very large positive integer to be determined from the lattice properties and  a partition function of the form:
\begin{eqnarray}
Z=\frac{1}{(VT)^{NN_1}}\prod_p (p^2)^{N_1}\exp[-E_02T]=X_N\exp[-E_02T],
\label{eqsw23}
\end{eqnarray}
where $N_1$ is an integer that can be both positive and negative we can apply basic thermodynamics relations as follows:
\begin{eqnarray}
&&S=UT+\ln(Z)=
-\frac{\partial \ln(X_N)}{\partial T}T-\frac{\partial[-2E_0T]}{\partial T}T+\ln(X_N)-2E_0T=
\nonumber\\
&&=-\frac{\partial \ln(X_N)}{\partial T}T+\ln(X_N)=NN_1+\ln X_N.
\label{res6647}
\end{eqnarray}
Since we expect for the entropy of a system at zero temperature to be zero or finite and the entropy in Eq. (\ref{res6647}) is infinite we conclude that one must set $N_1=0$ such that $X_N=1$ in order to obtain a correct result.  Thus this seems to suggest:

{\it A necessary and sufficient condition for the entropy of the system at zero temperature to be  zero or  finite and thus for the theory to be viable and complete even at the level of effective theory is that the partition function does not depend at all on the momenta and thus the corresponding field degrees  of freedom in the model are balanced in the  sense that the number of bosonic degrees of freedom is equal to the fermion ones}.

Note that $N_1=31$ for the standard model indicating that this theory is not complete even at the level of effective theory. Also it is evident that the supersymmetric extension of the standard model satisfies this criterion automatically.

\section{A useful path integral identity and  fermion operators}

We start from the fermion Lagrangian:
\begin{eqnarray}
&&{\cal L}=\bar{\Psi}A_0\Psi+\bar{\xi}B_0\xi+\bar{\chi}C_0\chi+
\nonumber\\
&&\frac{1}{M^4}\bar{\Psi}C^k\Psi\bar{\xi}D^k\xi+\frac{1}{M^4}\bar{\Psi}F^k\Psi\bar{\chi}E^k\chi+\frac{1}{M^4}\bar{\xi}G^k\xi\bar{\chi}H^k\chi.
\label{iftt4567}
\end{eqnarray}
Here $\Psi_i$, $\xi_i$ and $\chi_i$ are each a collection of $n$ fermions, $A_0$, $B_0$ and $C_0$ are operators with mass dimension $1$ and dimensions $4n\times 4n$ whereas
$C^k$, $D^k$, $F^k$, $E^k$, $G^k$, $H^k$ are operators in the same space and the index $k=1...m$. The repeated indices are summed over.

One can represent the Lagrangian in Eq. (\ref{iftt4567}) as a path integral over the fields $J_k$, $S_k$, $X_k$, $Y_k$ , $U_k$, $V_k$ ($k=1..m$) for the modified Lagrangian:
\begin{eqnarray}
&&{\cal L}_1=\bar{\Psi}A_0\Psi+\bar{\xi}B_0\xi+\bar{\chi}C_0\chi+
\nonumber\\
&&\frac{1}{M}\bar{\Psi}C^k\Psi J_k+\frac{1}{M}\bar{\xi}D^k\xi S_k+\frac{1}{M}\bar{\chi}E^k\chi Y_k+
\nonumber\\
&&\frac{1}{M}\bar{\Psi}F^k\Psi X_k+\frac{1}{M}\bar{\xi}G^k\xi U_k+\frac{1}{M}\bar{\chi}H^k\chi V_k+
\nonumber\\
&&M^2J_kS_k+M^2X_kY_k+M^2U_kV_k.
\label{newlagr554}
\end{eqnarray}

In terms of the partition function one can write:
\begin{eqnarray}
&&Z=\int d \bar{\Psi} d\Psi d\bar{\xi} d \xi d \bar{\chi} d\chi\exp[i\int d^4 x{\cal L}]=
\nonumber\\
&&\int d \bar{\Psi} d\Psi d\bar{\xi} d \xi d \bar{\chi} d\chi d J_k d S_k d X_k d Y_k d U_k d V_k \exp[i\int d^4x {\cal L}_1],
\label{part4554}
\end{eqnarray}
since the integrals over the fields $J_k$, $X_k$, $U_k$ lead to delta functions in terms of $S_k$, $Y_k$ and $V_k$ in the functional space.

Considering the structure of the Lagrangian ${\cal L}_1$ one can first integrate in the partition function over the fermions $\Psi$, $\xi$ and $\chi$ and their conjugates to obtain:
\begin{eqnarray}
&&Z=\int d J_k d S_k d X_k d Y_k d U_k d V_k \times
\nonumber\\
&&\det[A_0+\frac{C^kJ_k}{M}+\frac{F^kX_k}{M}]\times
\nonumber\\
&&\det[B_0+\frac{D^kS_k}{M}+\frac{G^kU_k}{M}]\times
\nonumber\\
&&\det[C_0+\frac{E^kY_k}{M}+\frac{H^kV_k}{M}]=
\nonumber\\
&&\int d J_k d S_k d X_k d Y_k d U_k d V_k \times
\nonumber\\
&&\det\Bigg[A_0[B_0+\frac{D^kS_k}{M}+\frac{G^kU_k}{M}][C_0+\frac{E^lY_l}{M}+\frac{H^lV_l}{M}]+
\nonumber\\
&&\frac{C^kJ_k}{M}[B_0+\frac{D^iS_i}{M}+\frac{G^iU_i}{M}][C_0+\frac{E^lY_l}{M}+\frac{H^lV_l}{M}]+
\nonumber\\
&&\frac{F^kX_k}{M}[B_0+\frac{D^iS_i}{M}+\frac{G^iU_i}{M}][C_0+\frac{E^lY_l}{M}+\frac{H^lV_l}{M}]\Bigg]\times
\nonumber\\
&&\exp[i\int d^4 x[M^2J_kS_k+M^2X_kY_k+M^2U_kV_k]].
\label{newexpr66577}
\end{eqnarray}

Knowing in the path integral formalism the determinant can be obtained as an integral over fermion variables we introduce $n$ new fermion fields $\tau_i$ and $\bar{\tau}_i$ such that:
\begin{eqnarray}
&&Z=\int d J_k d S_k d X_k d Y_k d U_k d V_k  \prod_{i=1}^n d \bar{\tau}_i d \tau_i\times
\nonumber\\
&&\exp\Bigg[i\frac{1}{M^2}\int d^4 x \bar{\tau}[A_0[B_0+\frac{D^kS_k}{M}+\frac{G^kU_k}{M}][C_0+\frac{E^lY_l}{M}+\frac{H^lV_l}{M}]+
\nonumber\\
&&\frac{C^kJ_k}{M}[B_0+\frac{D^iS_i}{M}+\frac{G^iU_i}{M}][C_0+\frac{E^lY_l}{M}+\frac{H^lV_l}{M}]+
\nonumber\\
&&\frac{F^kX_k}{M}[B_0+\frac{D^iS_i}{M}+\frac{G^iU_i}{M}][C_0+\frac{E^lY_l}{M}+\frac{H^lV_l}{M}]\tau\Bigg]\times
\nonumber\\
&&\exp[i\int d^4 x[M^2J_kS_k+M^2X_kY_k+M^2U_kV_k]].
\label{equv566}
\end{eqnarray}

Next step is to integrate over $J_k$ and $X_k$ noting that they appear only in liner terms and their integration leads to delta functions:
\begin{eqnarray}
&&Z=\int  d S_k  d Y_k d U_k d V_k  \prod_{i=1}^n d \bar{\tau}_i d \tau_i\times
\nonumber\\
&&\delta(M^2S_k+\frac{1}{M^3}\bar{\tau}C^k[B_0+\frac{D^iS_i}{M}+\frac{G^iU_i}{M}][C_0+\frac{E^lY_l}{M}+\frac{H^lV_l}{M}]\tau)\times
\nonumber\\
&&\delta(M^2Y_k+\frac{1}{M^3}\bar{\tau}F^k[B_0+\frac{D^iS_i}{M}+\frac{G^iU_i}{M}][C_0+\frac{E^lY_l}{M}+\frac{H^lV_l}{M}]\tau)\times
\nonumber\\
&&\exp[\int d^4 x\frac{1}{M^2}[\bar{\tau}[A_0[B_0+\frac{D^iS_i}{M}+\frac{G^iU_i}{M}][C_0+\frac{E^lY_l}{M}+\frac{H^lV_l}{M}]\tau+M^2U_kV_k]].
\label{del7756}
\end{eqnarray}

We shall consider the first delta function in Eq. (\ref{del7756}) as a function of $S_k$ and the second as a function of $Y_k$.
Knowing that $\delta(ax)=\frac{1}{a}\delta(x)$ we calculate:
\begin{eqnarray}
&&=\delta(M^2S_k+\frac{1}{M^3}\bar{\tau}C^k[B_0+\frac{D^iS_i}{M}+\frac{G^iU_i}{M}][C_0+\frac{E^lY_l}{M}+\frac{H^lV_l}{M}]\tau)\
\nonumber\\
&&\delta(S_k+\frac{\bar{\tau}\frac{C^k}{M^5}[B_0+\frac{D^iS_{i\neq k}}{M}+\frac{G^iU_i}{M}][C_0+\frac{E^lY_l}{M}+\frac{H^lV_l}{M}]\tau}{M^2+
\bar{\tau}\frac{C^kD^k}{M^4}(C_0+\frac{E^lY_l}{M}+\frac{H^lV_l}{M})\tau})\times
\nonumber\\
&&\prod_{k=1}^n[M^2+\bar{\tau}\frac{C^kD^k}{M^4}(C_0+\frac{E^lY_l}{M}+\frac{H^lV_l}{M})\tau]^{-1}.
\label{res5545}
\end{eqnarray}
Similarly:
\begin{eqnarray}
&&\delta(M^2Y_k+\frac{1}{M^3}\bar{\tau}[B_0+\frac{D^iS_i}{M}+\frac{G^iU_i}{M}][C_0+\frac{E^lY_l}{M}+\frac{H^lV_l}{M}]\tau)=
\nonumber\\
&&\delta(Y_k+\frac{\bar{\tau}\frac{F^k}{M^5}[B_0+\frac{D^iS_i}{M}+\frac{G^iU_i}{M}][C_0+\frac{E^lY_{l\neq k}}{M}+\frac{H^lV_l}{M}]\tau}
{M^2+\bar{\tau}\frac{F^k}{M^4}[B_0+\frac{D^lS_l}{M}+\frac{G^lU_l}{M}]E^k\tau}\times
\nonumber\\
&&\prod_{k=1}^n[M^2+\bar{\tau}\frac{F^k}{M^4}[B_0+\frac{D^lS_l}{M}+\frac{G^lU_l}{M}]E^k\tau]^{-1}.
\label{res442324}
\end{eqnarray}
If we introduce the expressions for $S_k$ and $Y_k$ in the exponent in Eq. (\ref{del7756}) we get terms of order $\frac{1}{M^8}$. Since we aim to work in order $\frac{1}{M^4}$ these terms should be dropped. Thus all contributions that contain $S^k $ and $Y^k$ in the exponent can be neglected.

Next we need to deal with the products that appear at the end of Eqs. (\ref{res5545}) and (\ref{res442324}). For that we first switch to the euclidean space by making the substitution
$t_E=-it$. Then we compute for example:
\begin{eqnarray}
&&\prod_{x_E}\prod_{k=1}^n[M^2+\bar{\tau}\frac{C^kD^k}{M^4}(C_0+\frac{E^lY_l}{M}+\frac{H^lV_l}{M})\tau]^{-1}\rightarrow
\nonumber\\
&&\prod_{x_E}\prod_{k=1}^n[1+\frac{1}{M^2}\bar{\tau}\frac{C^kD^k}{M^4}(C_0+\frac{E^lY_l}{M}+\frac{H^lV_l}{M})\tau]^{-1}=
\nonumber\\
&&\exp[-\sum_{x_E,k}\ln[1+\frac{1}{M^2}\bar{\tau}\frac{C^kD^k}{M^4}(C_0+\frac{E^lY_l}{M}+\frac{H^lV_l}{M})\tau]]\approx
\nonumber\\
&&\exp[-\sum_{x_E}\sum_k \frac{1}{M^2}\bar{\tau}\frac{C^kD^k}{M^4}(C_0+\frac{E^lY_l}{M}+\frac{H^lV_l}{M})\tau]=
\nonumber\\
&&\exp[i\int d^4 x\sum_k \bar{\tau}\frac{C^kD^k}{M^2}(C_0+\frac{E^lY_l}{M}+\frac{H^lV_l}{M})\tau]
\label{res66457}
\end{eqnarray}
Here we used:
\begin{eqnarray}
&&\frac{1}{M^4}\sum_{x_E}\rightarrow\int d^4x_E
\nonumber\\
&&\int d^4x_E\rightarrow-i\int d^4 x
\label{res5546}
\end{eqnarray}
and the sum $\sum_{x_E}$ is considered over all separate four space time coordinates. A similar expression is obtained  for the product at the end of Eq. (\ref{res442324}). We introduce both of them in the partition function in Eq. (\ref{del7756}) and apply the delta functions to get:
\begin{eqnarray}
&&Z=\int d U_k d V_k d \prod_{i=1}^n \bar{\tau}_i d \tau_i
\exp[i\int d^4 x[\bar{\tau}[A_0[B_0+\frac{G^kU_k}{M}][C_0+\frac{H^lV_l}{M}]\tau+
\nonumber\\
&&i\bar{\tau}\frac{C^kD^k}{M^2}(C_0+\frac{H^lV_l}{M})\tau+
i\bar{\tau}\frac{F^k}{M^2}(B_0+\frac{G^lU^l}{M})E^k\tau+iM^2U_kV_k]].
\label{res435627}
\end{eqnarray}
We integrate over $V_k$ to obtain a delta function in $U_k$ and apply the previous procedure working again in order $\frac{1}{M^4}$.  This leads to the following partition function:
\begin{eqnarray}
&&Z =\int d \bar{\Psi} d\Psi d\bar{\xi} d \xi d \bar{\chi} d\chi\exp[i\int d^4 x{\cal L}]\approx
\nonumber\\
&&\int \prod_{i=1}^n \bar{\tau}_i d \tau_i\times
\exp [i\int d^4 x\frac{1}{M^2}[\bar{\tau}A_0B_0C_0\tau+\bar{\tau}C^kD^kC_0\tau+\bar{\tau} F^kB_0E^k\tau+\bar{\tau}A_0G^kH^k\tau]]=
\nonumber\\
&&\det[\frac{i}{M^2}[A_0B_0C_0+C^kD^kC_0+F^kB_0E^k+A_0G^kH^k]].
\label{final66545}
\end{eqnarray}.

\section{Higher dimension operators}

The standard model is a powerful but incomplete theory from many points of view. Looking at the standard model from the point of view of the criterion stated in section II one would need additional $62$ real bosonic degrees of freedom that might be distributed among scalars and gauge bosons. However is it not our purpose here to determine a completion at higher scales but to evaluate what kind of operators one might introduce to make the effective theory consistent at or below the electroweak scale. It is understood that these new terms in the Lagrangian must be function of the known standard model degrees of freedom. In this perspective the only way to reestablish the correct expression for the partition function is to introduce operators that cancel the kinetic contribution of some of the fermions. This has been the whole purpose of the identity derived in section III. As an aside we note that it is assumed that by integrating out a particle its kinetic term will be disregarded from the partition function. It merely remains as a factor that cancels partially other particles integrated out at a larger scale.

We shall consider in what follows two possibilities: One is the standard model at the electroweak scale where the partition function $Z\approx (p^2)^{31}\times {\rm other\,factors}$; the second is below the electroweak scale where the $W$, $Z$, $h$ and $t$ has been integrated out and $Z\approx (p^2)^{30}\times {\rm other\,factors}$ (Note that the corresponding gauge fields are integrated out together with their ghosts).
Since in the identity in Eq. (\ref{final66545}) there is an uniformity of the dimension of the matrices involved  we cannot mix lepton states which are individual with quark states which are summed over color degrees of freedom (we need to preserve the invariance under the color group). The operators $A_0$, $B_0$, $C_0$, $C^k$, $D^k$, $F^k$, $E^k$, $G^k$, $H^k$ can thus act all of them either in the lepton or in the quark sector. We will exemplify this only for the up quarks as the extension to the the down quarks and leptons is straightforward.

Thus we make the substituiton:
\begin{eqnarray}
&&\Psi_i=(u_1 u_2 u_3)
\nonumber\\
&&\xi_i=(c_1 c_2 c_3 )
\nonumber\\
&&\chi_i=(t_1 t_2 t_3),
\label{res54664}
\end{eqnarray}
where the indices $1$, $2$, $3$ refer to the three colors.  The matrices $A_0$, $B_0$ are:
\begin{eqnarray}
&&A_0=(i\gamma^{\mu}\partial_{\mu}-m_{01})\times I_3
\nonumber\\
&&B_0=(i\gamma^{\mu}\partial_{\mu}-m_{02})\times I_3
\nonumber\\
&&C_0=(i\gamma^{\mu}\partial_{\mu}-m_{03})\times I_3
\label{not6657}
\end{eqnarray}
Here $I_3$ is a unit three dimensional matrix. Exactly the same setup can be applied identically to the bottom quarks and also to the leptons.

The next step is to compensate in the determinant in Eq. (\ref{final66545})  the action of the operator $A_0B_0C_0$.  For that we consider:
\begin{eqnarray}
&&C^k=x_1(i\gamma^{\mu}\partial_{\mu}+m_1)\times I_3
\nonumber\\
&&D^k=x_2(i\gamma^{\mu}\partial_{\mu}+m_2)\times I_3
\nonumber\\
&&F^k=x_3(i\gamma^{\mu}\partial_{\mu}+m_3)\times I_3
\nonumber\\
&&E^k=x_4(i\gamma^{\mu}\partial_{\mu}+m_4)\times I_3
\nonumber\\
&&G^k=x_5(i\gamma^{\mu}\partial_{\mu}+m_5)\times I_3
\nonumber\\
&&H^k=x_6(i\gamma^{\mu}\partial_{\mu}+m_6)\times I_3.
\label{res43435}
\end{eqnarray}
Note that one can modify these operators easily (by considering $i\gamma^{\mu}\partial-i\gamma^{\mu}\overleftarrow{\partial}$) such  that the final Lagrangian in Eq. (\ref{iftt4567}) is hermitian. Alternatively one can take the hermitian conjugate in the final result. For the sake of simplicity we shall not make this modification in what follows.

1) First we consider the theory slightly above the electroweak scale. Then one takes in Eq. (\ref{not6657}) all masses equal to zero ($m_{01}=m_{02}=m_{03}=0$). We shall state from the beginning that for this case the setup must be applied similarly to the bottom quarks, charged leptons and neutrinos. We want to cancel how many kinetic contributions is possible in the expression:
\begin{eqnarray}
A_0B_0C_0+C^kD^kC_0+F^kB_0E^k+A_0G^kH^k.
\label{res43556}
\end{eqnarray}
For that the following conditions must be fulfilled:
\begin{eqnarray}
&&x_1x_2+x_3x_4+x_5x_6+1=0
\nonumber\\
&&x_1x_2(m_1+m_2)+x_3x_4(m_3+m_4)+x_5x_6(m_5+m_6)=0.
\label{firstres6657}
\end{eqnarray}
Thus the first line cancel the $(i\gamma^{\mu}\partial_{\mu})^3$ contribution whereas the second line the $(i\gamma^{\mu}\partial_{\mu})^2$ one. The whole process applied to  all standard model fermions reduces the contribution of the kinetic term to that of a single generation. Then in the partition function this would produce  factors of $(p^2)^{15}$ instead of $(p^2)^{45}$ as it was initially the case. Since the bosons contribute factors of $\frac{1}{(p^2)^{14}}$ there is a mismatch of $p^2$.  Then one concludes that either there is an extra gauge boson in the theory or a complex scalar or one needs to consider a mass term for the neutrinos generated at a larger scale.  This is because the presence of a mass term for a fermion generically can produce the total cancelation of its kinetic contribution as we will illustrate next.

2) Below the electroweak scale it is assumed that operators in Eq. (\ref{res43435}) get modified by integrating out the Higgs boson and the other particles and thus act only in the quark sector. The conditions analogous to those in Eq. (\ref{firstres6657}) applied only in the quark sector become:
\begin{eqnarray}
&&x_1x_2+x_3x_4+x_5x_6+1=0
\nonumber\\
&&x_1x_2(m_1+m_2-m_{01})+x_3x_4(m_3+m_4-m_{02})+x_5x_6(m_5+m_6-m_{03})=0
\nonumber\\
&&x_1x_2(m_1m_2-m_1m_{01}-m_2m_{01})+x_3x_4(m_3m_4-m_2m_{02}-m_4m_{02})+x_5x_6(m_5m_6-m_5m_{03}4m_6m_{03})=0.
\label{res442333}
\end{eqnarray}
These conditions eliminate completely any momenta in the determinant in Eq. (\ref{final66545}) and thus the contribution of the kinetic terms of all quarks. Then the partition function will contain factors of $(p^2)^9$ coming only from the leptonic sector and $\frac{1}{(p^2)^9}$ coming from the bosonic sector. It is evident that these contributions cancel each other.

We offered here just a glimpse of what operators might exist without exhausting in any way the vast array of possibilities.

\section{Discussion and conclusions}

Before anything we need to examine what kind of particles and interaction might lead to operators of the kind suggested in Eq. (\ref{iftt4567}).
One possibility is to consider  a standard coupling of a pair of fermions with an extra Higgs boson $H_i$ (assume one for each generation) such that  the interactions for the Higgs bosons contains a term of the type:
\begin{eqnarray}
{\cal L}_h=aH_1H_2+bH_2H_2+cH_1H_3
\label{int6657}
\end{eqnarray}
where for simplicity we considered real scalar fields.
We thus start from the partial fermion term:
\begin{eqnarray}
{\cal L}_f=\bar{\Psi}(i\gamma^{\mu}\partial_{\mu}+xH_1)\Psi+...
\label{reqw23}
\end{eqnarray}
and apply the method introduced in \cite{Jora} for eliminating the scalar from the theory:
We consider the partition function:
\begin{eqnarray}
Z_f=\int d \bar{\Psi} d \Psi \exp[i\int d^4x[\bar{\Psi}(i\gamma^{\mu}\partial_{\mu}+xH_1)\Psi]=\det[i\gamma^{\mu}\partial_{\mu}+xH_1].
\label{part44432}
\end{eqnarray}
 We introduce another equivalent partition function $Z_1$ of the form:
\begin{eqnarray}
&&Z_{1f}=\int d \bar{\Psi} d \Psi d \bar{\tau} d \tau \exp[i\int d^4x\frac{1}{M^3}[\bar{\Psi}i(\gamma^{\mu}\partial_{\mu}+xH_1)\Psi\bar{\tau}\tau+v_1^3H_1\bar{\tau}\tau]]=
\nonumber\\
&&\det[i\gamma^{\mu}\partial_{\mu}+xH_1].
\label{res44355}
\end{eqnarray}
The second term in the exponent does not contribute because there must be a match between the degrees of freedom for $\Psi$ and $\tau$.
Now we make a change of variable from $\tau$ to $J_i$ where $J_1=\bar{\tau}\tau$ and $J_{j\neq1}$ are other fermion bilinears that sum up $8$ degrees of freedom:
\begin{eqnarray}
&&d \bar{\tau}  d\tau= dJ_1 \prod_j d J_{j\neq1}|\frac{dJ}{d\tau}|\approx
\nonumber\\
&&dJ_1  \prod_j d J_{j\neq1}\bar{\tau}_1\bar{\tau}_2...\tau_1\tau_2...\approx dJ_1 d J_{j\neq1}J_1^4.
\label{res64556}
\end{eqnarray}
The above relation holds up to some constant factor that is irrelevant (see \cite{Jora} for details).
Then $Z_1$ becomes:
\begin{eqnarray}
&&Z_{1f}=\int d \bar{\Psi} d \Psi dJ_1 \prod_j d J_{j\neq 1}J_1^4\exp[i\int d^4x\frac{1}{M^3}[\bar{\Psi}(i\gamma^{\mu}\partial_{\mu}+xH_1)\Psi J_1+v_1^3H_1J_1]]\rightarrow
\nonumber\\
&&\int d \bar{\Psi} d\Psi d \bar{\eta} d \eta  dJ_1 \prod_j d J_{j\neq 1}\exp[i\int d^4x[\frac{1}{M^3}[\bar{\Psi}(i\gamma^{\mu}\partial_{\mu}+xH_1)\Psi J_1+v_1^3H_1J_1]+\frac{1}{M_1^2}\bar{\eta}\eta J_1]]\rightarrow
\nonumber\\
&&\int d \bar{\Psi} d \Psi \prod_j d J_{j\neq 1}\delta(\bar{\Psi}(i\gamma^{\mu}\partial_{\mu}+xH_1)\Psi+v_1^3H_1)
\label{res41213}
\end{eqnarray}
Here we can drop the unwanted integrals $d J_{j\neq1}$ because they do not contribute to any interactions. One can take $M_1$ very large  and neglect also the term proportional to the factor $\frac{1}{M_1^2}$ in the exponent. The delta function should be solved in terms of $H_1$ and is equivalent to the substitution:
\begin{eqnarray}
H_1=\frac{\bar{\Psi}i\gamma^{\mu}\partial_{\mu}\Psi}{v^3}+...,
\label{res453667}
\end{eqnarray}
where the dots stand for further terms in the expansion.
This shows an example of  how four fermion interaction terms as those in Eq. (\ref{iftt4567}) might appear from simple interactions of the fermions with additional scalars in the standard model.

   In this work we considered a possible criterion for the consistency of a theory (effective or not) to be the momentum independency of the partition function  and examined some of its consequences. For that to happen either the number of fermion degrees of freedom must match the number of bosonic degrees of freedom or  the theory must contain higher dimensional operators  that cancel the extraneous kinetic contribution.
According to our criterion of consistency the UV completion of the standard model should contain additional $62$ real bosonic degrees of freedom. Knowing how the gauge bosons and scalars contribute to the partition function one may write:
\begin{eqnarray}
31=N_g+N_s,
\label{res5546}
\end{eqnarray}
where $N_g$ is the number of gauge bosons and $N_s$ is the number of complex scalars.  If one introduces right handed neutrinos this number becomes $34$.

We showed examples of  a new type of operators that can equilibrate the kinetic contribution in the partition function. In the absence of additional scalar degrees of freedom  these operators or versions of them are necessary for the theory to be complete as an effective theory but do not exhaust the list of possible higher dimensional terms that might appear in the Lagrangian. We also showed that these operators might come from new possible scalar states that interact with the fermions pairs. A more detailed discussion about possible new scalar contributions to the standard model will be made in a future work.

\section*{Acknowledgments} \vskip -.5cm

The work of R. J. was supported by a grant of the Ministry of National Education, CNCS-UEFISCDI, project number PN-II-ID-PCE-2012-4-0078.


\begin{thebibliography}{15}
\bibitem{Glashow} S. L. Glashow, Nuxl. Phys. {\bf 22} (4): 579-588 (1961).
\bibitem{Weinberg} S. Weinberg, Phys. Rev. Lett.  {\bf 19} (21):1264-1266 (1967).
\bibitem{Brout} F. Englert and R. Brout, Phys. Rev. Lett. {\bf 13} (16):321-323 (1964).
\bibitem{Higgs} P. W. Higgs, Phys. Rev. Lett.  {\bf 13}  (16): 508-509 (1964).
\bibitem{Salam} A. Salam, Elementary Particle Theory, Nobel Symposium No. 8, N. Svartholm (eds) (Almqvist and Wiksells, Stockholm 1968), p. 137.
\bibitem{Guralnik} G. S. Guralnik, C. R. Hagen and T. W. B. Kibble, Phys. Rev. Lett. {\bf 13}, 585 (1964).
\bibitem{Hooft} G. 't Hooft, Nucl. Phys. B {\bf 3}, 167 (1971).
\bibitem{Veltman} G. 't Hooft and M. J. G. Veltman, Nucl. Phys. B {\bf 44}, 189 (1972).
\bibitem{Peskin} M.E. Peskin and D. V. Schroeder, "An Introduction to Quantum Field Theory", Westview Press 1995.
\bibitem{Jora} A. H. Fariborz and R. Jora, arXiv:1509.05993 (2015).

\end{thebibliography}
\end{document}